\documentstyle[twocolumn,aps,epsf]{revtex}

\newcommand{\ewxy}[2]{\setlength{\epsfxsize}{#2}\epsfbox[10 60 640 570]{#1}}

\begin{document}

\draft        % to get PACS

\preprint{OUTP--98--52--P\\HD-THEP-98-31}
\title{String Breaking in Non-Abelian Gauge Theories with
Fundamental Matter Fields}
\author{Owe Philipsen$^1$ and Hartmut Wittig$^2$
\thanks{PPARC Advanced Fellow; member of UKQCD Collab.}
}
\address{$^1$Institut f\"ur Theoretische Physik, Philosophenweg~16,
D-69120 Heidelberg, Germany,\\
email: o.philipsen@thphys.uni-heidelberg.de}
%\author{Hartmut Wittig\thanks{PPARC Advanced Fellow}}
\address{$^2$Theoretical Physics, 1~Keble Road, Oxford OX1~3NP, UK,\\
email: h.wittig1@physics.oxford.ac.uk}

% \date{\today}

\maketitle

\begin{abstract}
%%HW
We present clear numerical evidence for string breaking in
three-dimensional SU(2) gauge theory with fundamental bosonic matter
through a mixing analysis between Wilson loops and meson operators
representing bound states of a static source and a dynamical
scalar. The breaking scale is calculated in the continuum limit. In
units of the lightest glueball we find $r_{\rm b}\,m_G\approx13.6$. The
implications of our results for QCD are discussed.
\end{abstract}

% Lattice gauge theory, General properties of QCD, Field theories in
% dimensions other than four

\narrowtext

%{\bf Introduction.}

It has been a long-standing problem in QCD to detect
``string-breaking'', i.e. the breakdown of linear confinement, by means
of computer simulations. String breaking is expected to occur when the
energy of a gauge string between static colour sources separated by a
distance~$r$ becomes as large as the energy required to produce a
light quark-antiquark pair, each of which is subsequently bound to the
static colour sources. Thus, for distances~$r$ below a typical scale
$r_{\rm b}$ the static quarks support a gauge string, resulting in a
linearly rising potential, whereas for $r> r_{\rm b}$ the ground state
potential is bounded by the energy of two static-light ``mesons'',
$V(r)\simeq E_{MM}$. The usual computational strategy is to extract
$V(r)$ from Wilson loops calculated in Monte Carlo simulations and to
look for signs that the linear rise in $V(r)$ turns into a constant
behaviour as the separation~$r$ is increased beyond the breaking
scale. Despite recent efforts
\cite{SESAM_pot96,UKQCD_lat97,CP-PACS_lat97}, hard evidence for string
breaking has not been presented so far.

It has been suggested that the apparent failure to detect string
breaking is due to insufficient overlap of Wilson loops with the
two-meson state \cite{guesken_lat97}. Indeed, a recent analysis of the
problem using strong coupling ideas \cite{itd98} suggests that string
breaking is a mixing phenomenon, involving both the string and the
two-meson state. Thus, in order to confirm the mixing picture, the
conventionally used Wilson loops have to be supplemented by explicit
two-meson operators. However, given the complexity and computational
cost of full QCD simulations, it will take a big effort to obtain an
unambiguous signal for string breaking, even if mixing is taken into
account.

In this letter we study the mixing between the string and a
two-``meson'' state in a simplified theory, the three-dimensional
SU(2) Higgs model. We present clear evidence for string breaking from
Monte Carlo simulations and essentially confirm the mixing scenario of
ref.~\cite{itd98}. Although the three-dimensional SU(2) Higgs model is
usually applied to physical contexts other than the strong
interactions, its phase diagram has a confinement region that shares a
number of features with QCD, such as a linearly rising potential of
static colour charges and its eventual screening through matter pair
creation.  As in QCD, the screening phenomenon has so far not been
observed in simulations employing Wilson loops only \cite{ilg97}.  
%%HWREV
In our previous studies of the model \cite{paper1,paper2,paper3} we
have investigated a possible mechanism how the confining properties of
the model are lost in the Higgs region through flux tube decay
\cite{paper2}. Furthermore, an attempt to estimate the screening
length by using non-local gauge-invariant operators has been described
in \cite{paper3}.

Bosonic matter fields imply an enormous simplification of the
computational effort, whilst preserving the underlying mechanism for
string breaking to occur. The presence of a weak or even moderately
strong scalar self-coupling, which has no analogue in QCD, is not
expected to change the dynamics of gauge and scalar fields so
dramatically that the interpretation of results in a QCD context is no
longer possible \cite{paper1,paper2}.

The study of the model in three dimensions also offers a number of
advantages: first one can simulate relatively large volumes, so that
large separations of colour charges can be probed efficiently without
suffering from significant finite-volume effects. Second, due to
superrenormalisability, the curves of constant physics can be mapped
out exactly using two-loop perturbation theory \cite{lai95}.
Furthermore, the bare gauge coupling $g^2$ is a dimensionful quantity,
which sets the scale in three-dimensional gauge theories.

We work with the lattice action defined by
\begin{eqnarray}
& S[U,\phi] &~= \beta_G\sum_p\left(1-\textstyle\frac{1}{2}{\rm Tr\,}
   U_p\right)    \\
&+&\hspace{-2mm}\sum_x\Big\{-\beta_H\sum_{\mu=1}^3\textstyle\frac{1}{2}
{\rm Tr\,}\big(\phi^{\dagger}(x)U_\mu(x)
\phi(x+\hat{\mu})\big)  \nonumber\\
&+&\hspace{-2mm}\textstyle\frac{1}{2}{\rm
   Tr\,}\big(\phi^{\dagger}(x)\phi(x)\big) 
+\beta_R\left[\textstyle\frac{1}{2}
{\rm Tr\,}\big(\phi^{\dagger}(x)\phi(x)\big)-1\right]^2 
\Big\}, \nonumber
\end{eqnarray}
where $U_\mu(x)\in\rm SU(2)$ is the link variable, $U_p$ denotes the
plaquette, and $\phi(x)$ is the scalar field. The bare parameters
$\beta_G\equiv4/(ag^2), \beta_H$ and $\beta_R$ are the inverse gauge
coupling, the scalar hopping parameter, and scalar self-coupling,
respectively. In our Monte Carlo simulations we update the gauge
fields using a combination of the standard heatbath and
over-relaxation algorithms for SU(2)\,\cite{fabhaan,kenpen}. The
scalar fields are updated using the algorithm described
in~\cite{bunk_lat94}. Further details about our simulation procedure
can be found in ref.~\cite{paper1}.

%%HW Wilson loop and strings
The standard operator used to compute the static potential is the
Wilson loop, i.e. the correlation of a string of length $r$ over a
time interval $t$. Throughout this work we have computed Wilson loops
of area $r\times t$ for ``on-axis'' orientations in the spatial
directions, so that all separations are integer multiples of the
lattice spacing $a$:
%%HW shortened
% Thus, our Wilson loops of area $r\times t$ are defined as
%
\begin{eqnarray}
& G_{SS}(r,t)&~ = \Big\langle{\rm Tr\,}\Big\{
% U(x,x+r\hat{\j})\,U(x+r\hat{\j},x+r\hat{\j}+t\hat{3}) \nonumber\\
 U(0,r\hat{\j})\,U(r\hat{\j},r\hat{\j}+t\hat{3}) \nonumber\\
&\times&\hspace{-5mm}
% U^\dagger(x+t\hat{3},x+r\hat{\j}+t\hat{3})\,U^\dagger(x,x+t\hat{3}) 
 U^\dagger(t\hat{3},r\hat{\j}+t\hat{3})\,U^\dagger(0,t\hat{3}) 
 \Big\}\Big\rangle,
\label{eq_wilson}
\end{eqnarray}
where $U(x,y)$ is a shorthand notation for the straight line
of links connecting the sites $x$ and $y$. A non-local,
%%OP gauge invariant changed
gauge-invariant operator describing the bound state of a static colour
source and a scalar field in three dimensions has been studied in
\cite{paper3}, 
viz.
\begin{equation}
G_M(t) = \left\langle\textstyle\frac{1}{2}\,{\rm Tr\,}\left\{
 \phi^\dagger(t\hat{3})\,U^\dagger(0,t\hat{3})\,\phi(0)
 \right\}\right\rangle.
\end{equation}
This motivates our definition of an operator which projects onto two
of these bound states separated by a distance~$r$
\begin{eqnarray}
& G_{MM}(r,t)&~ = \Big\langle\textstyle\frac{1}{4}\,{\rm Tr\,}\Big\{
 (1-\sigma_3)\phi^\dagger(t\hat{3})\,U^\dagger(0,t\hat{3})\,\phi(0)\Big\}
\label{eq_twomeson}  \\
\hspace{-9mm}&\times&\hspace{-5mm}{\rm Tr\,}\Big\{(1-\sigma_3)
 \phi^\dagger(r\hat{\j})\,U(r\hat{\j},r\hat{\j}+t\hat{3})\,
 \phi(r\hat{\j}+t\hat{3})\Big\}
 \Big\rangle. \nonumber
\end{eqnarray}
Finally, we consider operators which describe transitions between a
string and a two-meson state and vice-versa:
\begin{eqnarray}
& G_{SM}(r,t)&~ = \Big\langle\textstyle\frac{1}{2}\,{\rm Tr\,}\Big\{
 \phi^\dagger(t\hat{3})\,U^\dagger(0,t\hat{3})\,U(0,r\hat{\j})
 \nonumber\\
\hspace{-9mm}&\times&\hspace{-5mm}
 U(r\hat{\j},r\hat{\j}+t\hat{3})\,\phi(r\hat{\j}+t\hat{3})\Big\}
 \Big\rangle \\
& G_{MS}(r,t)&~ = \Big\langle\textstyle\frac{1}{2}\,{\rm Tr\,}\Big\{
 \phi^\dagger(r\hat{\j})\,U(r\hat{\j},r\hat{\j}+t\hat{3})
 \nonumber\\
\hspace{-9mm}&\times&\hspace{-5mm}
 U^\dagger(t\hat{3},r\hat{\j}+t\hat{3})\,U^\dagger(0,t\hat{3})\,\phi(0)
 \Big\}\Big\rangle.
\end{eqnarray}
In order to study the static potential and the mixing between gauge
string and two-meson states, we construct the following matrix
correlator
\begin{equation}
 G(r,t) = \left(\begin{array}{ll}
	G_{SS}(r,t) & G_{SM}(r,t) \\
	G_{MS}(r,t) & G_{MM}(r,t)
		\end{array}\right).
\label{eq_matcor}
\end{equation}
The operator $G_M(t)$ serves to extract the energy $E_M$ of a single
bound state of a static source and a scalar field, which helps to
identify the two-meson state, since one naively expects
$E_{MM}\simeq2E_M$.

In order to enhance the signal of all our operators, we have
constructed smeared gauge and scalar fields. Following the standard
algorithm in~\cite{albanese} we constructed ``fuzzed'' link variables
of length~$a$. Fuzzed scalar fields were constructed using a suitable
adaptation of the blocking algorithm used
in~\cite{paper1,paper2,paper3}. In particular, eq.~(3) of
ref.~\cite{paper2} carries over literally after replacing the blocked
links by their fuzzed counterparts. We constructed spatial Wilson
lines for three different link fuzzing levels, whereas five fuzzing
levels were used for scalar fields, so that we obtained the correlator
$G(r,t)$ as an $8\times8$ matrix.

%%OP Our -> A
A procedure of diagonalising $G(r,t)$ following a variational method
has been described in detail in~\cite{paper1,mic}. 
Here we only wish to point out that the diagonalisation of the
$8\times8$ correlation matrix $G_{ik}$ yields eigenvectors $\Phi_i$
from which the correlation function of the (approximate) eigenstates
of the Hamiltonian can be calculated,
\begin{equation}
  \Gamma_i(t) = \langle \Phi_i(t) \Phi_i(0) \rangle = 
\sum_{j,k=1}^8\,a_{ij}a_{ik}\,G_{jk}(t).
\label{eq_eigenstate}
\end{equation}
The coefficients $a_{ik}$ quantify the overlap of each individual
correlator $G_{ik}$ with the correlators of the mass eigenstates, $\Gamma_i$. 
The diagonalisation procedure allows the computation of the ground state
energy, the first few excitations, and, most crucially, their
composition in terms of the original operator basis. The coefficients
$a_{ik}$ play a central role for the interpretation of string breaking
as a mixing phenomenon.

Our choice of bare parameters followed closely our earlier
work~\cite{paper1,paper2,paper3}. In particular, we approach the
continuum limit for a fixed ratio $\beta_G\,\beta_R/\beta_H^2 =
0.0239$, corresponding to a fixed continuum scalar self-coupling.  We
selected three values for the bare gauge coupling, i.e. $\beta_G=5.0,
7.0, 9.0$, and the hopping parameter $\beta_H$ was always chosen such
that the model was in the confinement region of the phase
diagram. Lattice sizes were chosen such that $L/\beta_G\simeq5.7$. At
$\beta_G=5.0$ we also considered $L/\beta_G=7.2$ in order to check for
finite-size effects. The maximum separation of colour charges was
$r_{\rm{max}}=L/2$ in all simulations. At $\beta_G=5.0$ we used 30
iterations in the fuzzing algorithm, which was sufficient to observe
saturation in the projection onto the ground state. The link/staple
mixing ratio was always set to two. For the runs at smaller lattice
spacing, the maximum number of fuzzing iterations was scaled by the
respective ratio of $\beta_G$-values. The correlators $G(r,t)$ and
$G_M(t)$ were measured after every full update, and 50 individual
measurements were collected in bins for post-processing. Typically we
performed 1500 measurements (30 bins) and 3000 measurements for our
most accurate determinations. Statistical errors were estimated using
a jackknife procedure.

\begin{figure}
\vspace{-2.0cm}
\ewxy{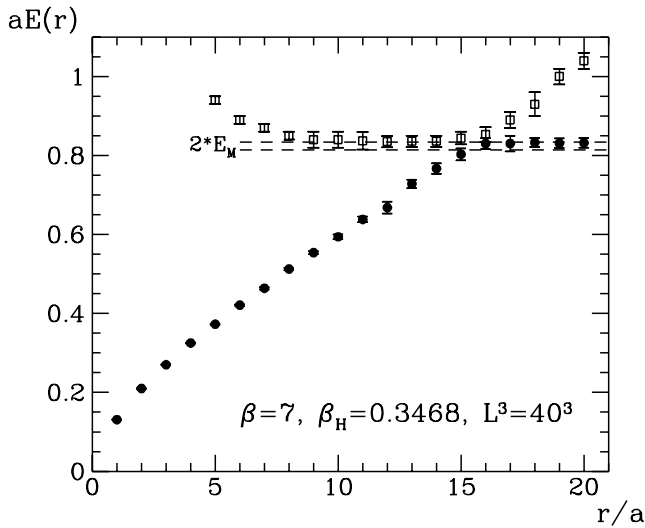}{130mm}
\vspace{-1.5cm}
\caption{The energies of the ground state and the first excited state
for $\beta_G=7.0, \beta_H=0.3468$ on $40^3$. The dashed lines indicate
the location of twice the energy of the single meson state, as
extracted from $G_M(t)$.
\label{fig_pot_b7}}
\end{figure}

In Fig.~\ref{fig_pot_b7} we show the energies of the two lowest
eigenstates as a function of $r/a$. The ground state shows the
familiar linear rise for small separations. For $r/a\ge15$ however,
the linear rise saturates, and a comparison with the energy extracted
for the single meson state shows that $E(r)\simeq2E_M$ in this
region. The qualitative and quantitative behaviour of the ground state
energy at large separations is entirely consistent with the string
breaking picture in the sense that the ground state energy is bounded
by that of the two-meson state. Thus, we take $r_{\rm b}\simeq15a$ as a
rough estimate at this point in parameter space. The energy of the
first excited state is constant for $r<r_{\rm b}$ and consistent with that
of a two-meson state. A significant $r$-dependence in this region is
only observed at very small separations. There are a number of
possible explanations, such as interactions between the two mesons,
mixing with the excited potential,
%%OP shortened, we do not know whether it's a fact
%but also the fact that the separation may be too small for
or that the separation may be too small for 
a two-meson state to exist. By contrast, for $r>r_{\rm b}$, the energies of
the first excited state rise linearly and appear to be the
continuation of the ground state potential below the breaking scale.

\begin{figure}
\vspace{-0.8cm}
\ewxy{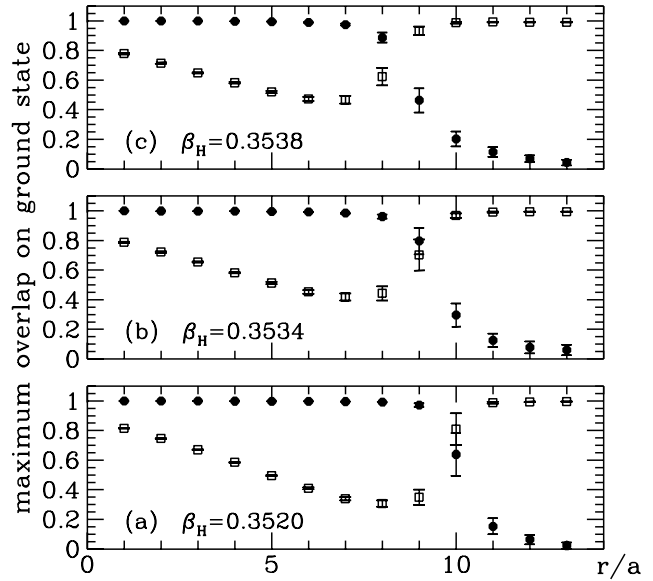}{130mm}
\vspace{-1.7cm}
\caption{The maximum projections on the ground state of the string
($a_{1,SS}$, full circles) and the two-meson operator
($a_{1,MM}$, open squares) for several values of the hopping parameter
at $\beta_G=5.0$ on $26^3$.
\label{fig_proj_b5}}
\end{figure}

The interpretation of the $r$-dependence of the spectrum as clear
evidence for string breaking is corroborated by the analysis of the
composition of the energy eigenstates in terms of the original
operator basis. In Fig. \ref{fig_proj_b5}~(a) we plot the maximum
projection of the string and two-meson operator onto the ground
state. 
%%HW Wilson loop --> string
At small $r/a$ it is obvious that the ground state is dominated by the
string operator, consistent with the picture of static quarks bound by
a flux tube, whereas the projection of the two-meson operator onto the
ground state is always significantly lower, but clearly non-vanishing.
%%HWREV
As mentioned above, the admixture of $G_{MM}$ to the ground state may
have physical reasons, but it could also arise through overlapping
fuzzed links in $G_{MM}$, which effectively create a loop of links.
Operator artefacts of this kind were already encountered in
\cite{paper2}. Thus, the most likely interpretation of the overlaps at
small $r/a$ is a combination of physical effects and operator
artefacts. With increasing $r/a$ $G_{MM}$ almost fully projects onto
the two-meson state. For $r\simeq r_{\rm b}$ however, there are clear
indications for mixing between the flux tube and the two-meson state,
with comparable projections of both operator types. For $r>r_{\rm b}$
the ground state receives almost exclusively contributions from the
two-meson operator, and it is now the first excited state (not shown),
whose dominant operator content comes from
%%HW Wilson loop --> string
the string. We conclude that there is clear evidence for the
crossing of energy levels associated with gauge strings and two-meson
operators at $r_{\rm b}$, so that the flux tube is energetically
disfavoured for $r>r_{\rm b}$.

At both $\beta_G=5.0, \beta_H=0.3520$ and $\beta_G=7.0,
\beta_H=0.3468$ we observe a fairly sharp crossover and thus a
relatively narrow mixing region. In order to explore the mixing
phenomenon for smaller scalar masses, we have simulated a sequence of
$\beta_H$-values at $\beta_G=5.0$ fixed, which all satisfy the
constraint $\beta_G\,\beta_R/\beta_H^2 = 0.0239$, i.e.~correspond to
the same continuum scalar coupling. For smaller scalar mass (larger
$\beta_H$) one expects $r_{\rm b}$ to occur at smaller separations,
since less energy is required to support a two-meson state. Also,
according to \cite{itd98} the width of the mixing region becomes
larger for smaller scalar mass, so that the crossover is expected to
widen as $\beta_H$ is increased. Our plots of the maximum projections
as a function of $\beta_H$ displayed in
Fig. \ref{fig_proj_b5}~(a)--(c) show that both is indeed the case.
The conclusions of \cite{itd98}, which have been obtained for SU(N)
gauge theories with fermions, are thus confirmed qualitatively in our
simplified model.

By comparing the spectrum for different $\beta_H$ at $\beta_G=5.0$
fixed, we observe that energies extracted from eigenstates with
%%HW Wilson loop --> gauge string
dominant admixtures of gauge strings remain largely unaffected by the
value of $\beta_H$, and only the energies of states with a dominant
overlap from two-meson operators show a significant dependence on
$\beta_H$. This is in line with our earlier observation that scalar
and gauge sectors decouple approximately \cite{paper1,paper2}. The
%%HW Wilson loop --> string operator
consequence of this in the present context is that the string operator
predominantly projects on the state with unbroken flux tube for {\it
%%HW rephrasing of Wilson loop
all\/} spatial separations outside the mixing region. Hence,
%%HW Wilson loop in the right context
Wilson loops exhibit an
area law for correlation distances $t/a$ up to which the signal can be
followed, even if $r/a$ is already beyond $r_{\rm b}$.  This offers a
possible explanation for the failure to detect string breaking in full
QCD simulations using only Wilson loops. Namely, since the only
%%HW Wilson loop --> string, string --> flux tube
significant overlap of a string onto the broken flux tube occurs
inside the mixing region and vanishes again when $r/a$ is increased
further, the saturation of the energy can only be observed if the
mixing region is broader than, say, two lattice spacings. Furthermore,
the width of the mixing region is sensitive to the mass of the matter
considered. Therefore, in computations of Wilson loops in full QCD
with relatively heavy dynamical quarks, the mixing region may be too
narrow. If this scenario holds, then the problem can only be overcome
by employing a variational approach of the type discussed in this
letter.

We now consider the scaling behaviour of the breaking scale $r_{\rm b}$. As
%%HW Wilson loop --> string
a definition of $r_{\rm b}$ we use the point where the maximum overlaps of
string and two-meson operators onto the ground state are equal, i.e.
\begin{equation}
  \Delta\equiv a_{1,SS}-a_{1,MM}=0,
\end{equation}
where we have borrowed from the notation in eqs.~(\ref{eq_wilson}),
(\ref{eq_twomeson}) and (\ref{eq_eigenstate}). An estimate of $r_{\rm b}/a$
is obtained through a local, linear interpolation of $\Delta$ to the
point where it vanishes. This procedure becomes exact in the continuum
limit. Within our accuracy we assign an error of at most one lattice
spacing to the estimate of $r_{\rm b}/a$.  Our results for all lattice
sizes and parameter values are shown in Table~\ref{tab_rbreak}. We
note again that $r_{\rm b}/a$ decreases as the scalar mass is
decreased. Furthermore, by comparing the estimates for $r_{\rm b}/a$ on
$L/a=26$ and~36 at $\beta_G=5.0$ we conclude that finite-size effects
are practically absent in our simulations.  In the last column $r_{\rm b}$
is given in units of the continuum gauge coupling constant $g^2$.
Within errors we observe scaling of the estimates for $r_{\rm b}\,g^2$ from
runs $c,\,e$ and $f$, whose bare parameters are related through lines
of constant physics in the continuum. For a controlled continuum
extrapolation, a further value at larger $\beta_G$ would be desirable,
so we choose to quote as our final result
\begin{equation}
r_{\rm b}\,g^2 \approx 8.5.
\end{equation}
%
%%HW subscript 0++ omitted in the following
This number can be combined with results for the mass spectrum in the
continuum limit at the same physical couplings \cite{paper1}. Taking
$m_{S}/g^2=0.839(15)$ for the lightest scalar bound state and
$m_{G}/g^2=1.60(4)$ for the lightest scalar glueball
\cite{paper1}, we obtain
\begin{equation}
r_{\rm b}\,m_{G} \approx 13.6,\;
m_{S}/m_{G}=0.524(16) 
\end{equation}
%
%%HWREV
To summarise, our results clearly show that string breaking occurs in
non-Abelian gauge theories with matter fields. The main evidence is
the existence of a mixing region, where both the string and two-meson
operators have a significant overlap onto the ground state. We have
also shown that Wilson loops are not suitable to compute the ground
state potential for $r>r_{\rm b}$. Estimates of $r_{\rm b}$ are also
of phenomenological interest. For instance, $r_{\rm b}$ is expected to
be related to the scale parameter of QCD through light-quark
constituent masses \cite{rosner96}. Also, $r_{\rm b}$ is important for
the understanding and description of fragmentation processes
\cite{lund83}. An extension of our analysis to QCD is therefore highly
desirable.
\begin{table}
\caption{Estimates for the breaking scale $r_{\rm b}$.
\label{tab_rbreak}
}
\begin{tabular}{ccccrc}
Run & $\beta_G$ & $\beta_H$ & $L/a$ & \multicolumn{1}{c}{$r_{\rm b}/a$}
    & $r_{\rm b}\,g^2$ \\
\tableline
a & 5.0 & 0.3538 & 26 &  $8.2\pm0.5$ & $6.56\pm0.38$ \\
b & 5.0 & 0.3534 & 26 &  $8.9\pm1.0$ & $7.12\pm0.79$ \\
c & 5.0 & 0.3520 & 26 &  $9.8\pm0.7$ & $7.88\pm0.59$ \\
d & 5.0 & 0.3520 & 36 &  $9.9\pm0.8$ & $7.90\pm0.64$ \\
e & 7.0 & 0.3468 & 40 & $14.8\pm1.0$ & $8.47\pm0.57$ \\
f & 9.0 & 0.3438 & 52 & $19.1\pm1.0$ & $8.50\pm0.44$ \\
\end{tabular}
\end{table}
%%%%%%%%%%%%%%%%%%%%%%%%%%%%%%%%%%%%%%%%%%%%%%%%%%%%%%%%%%%%%%%%%%%%%%%
% \clearpage
%\vskip 2mm
%\noindent {\bf Acknowledgments.}
We thank M. L\"uscher and M. Teper for useful discussions.
The computations have been performed on a NEC-SX4/32 at the HLRS
Stuttgart and the SGI Origin 2000 at University of Wales, Swansea. We
thank these institutions and the UKQCD Collaboration for their support.
\vspace*{-1.0cm}
%%%%%%%%%%%%%%%%%%%%%%%%%%%%%%%%%%%%%%%%%%%%%%%%%%%%%%%%%%%%%%%%%%%%%%%

%%%%%%%%%%%%%%%%%%%%%%%%%%%%%%%%%%%%%%%%%%%%%%%%%%%%%%%%%%%%%%%%%%%%%%%


\begin{references}

\vspace*{-1.5cm}
\bibitem{SESAM_pot96}
SESAM Collaboration (U. Gl\"assner et al.),
Phys. Lett. B383 (1996) 98.

\bibitem{UKQCD_lat97}
UKQCD Collaboration (M. Talevi),
Nucl. Phys. B (Proc. Suppl.) 63 (1998) 227.

\bibitem{CP-PACS_lat97}
CP-PACS Collaboration (S. Aoki et al.),
Nucl. Phys. B (Proc. Suppl.) 63 (1998) 221.

\bibitem{guesken_lat97}
S. G\"usken,
Nucl. Phys. B (Proc. Suppl.) 63 (1998) 16.

\bibitem{itd98}
I.T. Drummond, DAMTP-98-35, hep-lat/9805012.

\bibitem{ilg97}
M. G\"urtler, E.M. Ilgenfritz, J. Kripfganz, H. Perlt and A. Schiller,
Nucl.~Phys.~B483 (1997) 383.

\bibitem{lai95}
M.~Laine, Nucl.~Phys.~B451 (1995) 484.

\bibitem{paper1}
O. Philipsen, M. Teper and H. Wittig,
Nucl. Phys. B469 (1996) 445.

\bibitem{paper2}
O. Philipsen, M. Teper and H. Wittig,
Nucl. Phys.~B528 (1998) 379.

\bibitem{paper3}
M. Laine and O. Philipsen,
Nucl. Phys.~B523 (1998) 267.

\bibitem{fabhaan}
K. Fabricius and O. Haan,
Phys.~Lett. B143 (1984) 459.
 
\bibitem{kenpen}
A.D. Kennedy and B.J. Pendleton,
Phys.~Lett. B156 (1985) 393.
 
\bibitem{bunk_lat94}
B. Bunk,
Nucl.~Phys.~B (Proc. Suppl.) 42 (1995) 566.

\bibitem{albanese}
M.~Albanese et al.,
Phys.~Lett. B192 (1987) 163; Phys.~Lett. B197 (1987) 400.

\bibitem{mic}
L.A. Griffiths, C. Michael and P.E.L. Rakow, Phys.~Lett. B129 (1983) 351.

\bibitem{rosner96}
J.L. Rosner,
Phys. Lett. B385 (1996) 293.

\bibitem{lund83}
B. Andersson, G. Gustafson, G. Ingelman, T. Sjostrand,
Phys. Rep. 97 (1983) 31.

\end{references}
\end{document}